\documentclass[final,5p,times,twocolumn,authoryear]{elsarticle}
\usepackage{amsmath,amssymb}
\newtheorem{prop}{Proposition}
\newtheorem{defn}{Definition}
\usepackage{threeparttable}
\begin{document}
\begin{frontmatter}
\title{Buy-Online-and-Pick-up-in-Store in Omnichannel Retailing}
\author{Yasuyuki Kusuda\corref{cor1}}
\cortext[cor1]{
\emph{Email address:}\ \texttt{kusuda@n-fukushi.ac.jp} \\
\hspace{12pt} \emph{URL:}\ \texttt{https://handy.n-fukushi.ac.jp/pub/kusuda/index.html}
%\ead{kusuda@n-fukushi.ac.jp}
%\ead[url]{https://handy.n-fukushi.ac.jp/pub/kusuda/index.html}
}
\address{Faculty of Economics, Nihon Fukushi University, Aichi 477-0031, Japan}
\begin{abstract}
In this paper, we extend the model of \citet{Gao_Su2016} and consider an omnichannel strategy in which inventory can be replenished when a retailer sells only in physical stores.
With ``buy-online-and-pick-up-in-store'' (BOPS) having been introduced, consumers can choose to buy directly online, buy from a retailer using BOPS, or go directly to a store to make purchases without using BOPS.
The retailer is able to select the inventory level to maximize the probability of inventory availability at the store.
Furthermore, the retailer can incur an additional cost to reduce the BOPS ordering lead time, which results in a lowered hassle cost for consumers who use BOPS.
In conclusion, we found that there are two types of equilibrium: that in which all consumers go directly to the store without using BOPS and that in which all consumers use BOPS.
\end{abstract}
\end{frontmatter}
\section{Introduction}
As the use of the Internet has expanded in recent years, sales options called ``omnichannels'' have attracted more attention. 
Omnichannels integrate existing channels (\emph{online} and \emph{brick-and-mortar stores}), create seamless links between those channels, and provide consumers with information on goods and inventory availability. They are intended to promote purchases and expand retailer market coverage.
\par
One omnichannel method is ``buy-online-and-pick-up-in-store'' (BOPS).
BOPS is a fulfillment initiative that enables consumers to order goods online and pick them up at the nearest store.
By using BOPS, consumers can eliminate shipping costs and waiting times.
Also with BOPS, retailers can induce consumers to make additional in-store purchases when they come to stores to pick up items, thereby increasing sales.
For example, the Japanese convenience store chain \emph{Seven-Eleven Japan} developed an omnichannel service \emph{Omni 7} that offers BOPS to consumers.
As of July 2019, consumers who choose this option can avoid shipping fees.
\par
\citet{Gao_Su2016} focused on the BOPS information effect and showed that attracting consumers to stores in a rational expectations (RE) equilibrium is optimal.
In their view, consumers using BOPS have full access to product availability information, and they know that immediate replenishment of inventory is not possible when products are unavailable at stores.
They maintained that ``a useful by-product of BOPS is inventory availability information'' \citep{Gao_Su2016}.
Moreover, they showed that BOPS may not be profitable for all products, particularly for those that sell well in stores.
However, their model is not appropriate for cases in which stores can immediately replenish items that can be ordered through BOPS.
In this case, BOPS would not provide online information to consumers about inventory availability. This strategy has different implications for retailers.
\par
In this paper, we extend the model of \citet{Gao_Su2016} and consider an omnichannel strategy in which retailers can replenish inventory when they only sell items in physical stores.
Our research objective is to establish an alternative stylized model for BOPS in omnichannel retailing that is more realistic.
Notably, we aim to develop the model to reflect the convenience store industry with an omnichannel initiative and clarify consumer behaviors.
In this research, we use analytic models to investigate the hassle costs of consumers using different channels.
The significance of this paper is to demonstrate, in one model, the causality between consumer behaviors, hassle costs, delivery costs, and optimal omnichannel strategies.
In the model, we examine three kinds of consumer behaviors: purchasing items directly online; purchasing from a retailer using BOPS; and visiting a store in person to make purchases without using BOPS.
A retailer can select a level of store inventory at which consumers anticipate that inventory will be available.
However, in this case, BOPS would not provide complete information to consumers about inventory levels since immediate replenishment is possible. This is an essential contrast to \citet{Gao_Su2016}.
Furthermore, the retailer can incur an additional cost to reduce the lead time for BOPS ordering, which results in a lowered hassle cost (i.e., waiting time) for consumers using BOPS.
In conclusion, we found that there are two types of equilibrium: that in which all consumers go directly to a store to purchase items without using BOPS and that in which all consumers use BOPS to purchase items.
\par
BOPS is closely related to other topics in omnichannel research: inventory management, information effects, cross-selling, product returns, and others.
\citet{Bell_Gallino_Moreno2014} is one example of an excellent survey within omnichannel studies\footnote{
See also \citet{Brynjolfsson_Hu_Rahman2013}, \citet{Rigby2011}, \citet{Rosenblum_Kilcourse2013} and \citet{Hubner_Holzapfel_Kuhn2016}.
}.
They adopted ``a customer-focused perspective'' for an omnichannel strategy, asking two simple questions: (1) How will customers get the information they need to facilitate their purchase decisions? and (2) How will transactions be fulfilled?
To address these questions, they used a matrix with four quadrants:
\emph{offline-information and pickup},
\emph{online-information and pickup},
\emph{offline-information and delivery},
and
\emph{online-information and delivery}.
These correspond to traditional retail, showrooming, BOPS, and (pure) e-commerce, respectively.
Focusing on fulfillment, they enumerated the advantages and disadvantages of the in-store pickup and delivery options---a significant theme of this paper.
When orders are fulfilled in stores, vast on-site inventories may need to be maintained, meaning that BOPS may be associated with inventory management problems.
Consumers thus anticipate current inventory levels at stores to determine whether to visit the store to purchase particular products.
\citet{Su_Zhang2009} studied two types of retailer strategies---commitment and availability---to improve profit.
In other words, for the retailer, a challenge of BOPS is knowing how much information to share with consumers about current inventory levels.
\citet{Allon_Bassamboo2011} studied a retail operations model in which the retailer is strategic about providing information.
Meanwhile, one advantage of drawing consumers to stores is the cross-selling effect, or the possibility that customers will make additional purchases during the store visit.
In this case, a retailer would have the potential to realize cross-selling profits by offering BOPS, as consumers would have to visit the store to pick up their ordered goods.
\citet{Li_Sun_Wilcox2005} presented empirical results on the cross-selling effect, and \citet{Gallino_Moreno2014} examined BOPS impact on it\footnote{See also \citet{Netessine_Savin_Xiao2006}}.
The cross-selling effect is an issue only when a retailer has physical as well as online stores, since cross-selling opportunities arise only when customers purchase items at physical stores.
Some studies, including \citet{Gao_Su2016}, have pointed out that cross-selling opportunities can be profitable advantages for physical stores as opposed to online ones.
In this model, we consider a retailer who is using an omnichannel strategy alongside EC venders such as Amazon.com.
These retailers are using only a single channel, and thus, we do not have to consider cross-selling effects.
We do not discuss product return (\citet{Ofek_Katona_Sarvary2011}) or drop-shipping (\citet{Netessine_Rudi2006}) in order to focus on BOPS as the optimal fulfillment strategy.
We also ignore the price-matching problem, assuming price homogeneity between online and in-store channels.
Instead, in this paper, we highlight the hassle costs associated with consumer choices.  
\par
The next section describes the Japanese convenience store as a typical example of an omnichannel.
In Section 3, we formulate the model for the BOPS option. In Section 4, we discuss the results derived from the model. 
We then demonstrate a numerical example in Section 5, extend the model in Section 6, discuss policy suggestions in Section 7, and conclude the paper in Section 8.
\section{An Example: Omnichannels in Japanese Convenience Stores}
The omnichannel operated by Japanese convenience stores is an ideal system for the subject of our research.
Let us consider Omni 7,the omnichannel for Seven-Eleven Japan.
Omni 7 was introduced in November of 2015 as an omnichannel system combining online and physical stores. It allows the consumer to choose between home delivery and in-store pick up of purchased items\footnote{
For details in English, see the following website. \\
https://siklejapanpoint.work/2019/03/16/20-reduction-in-online-shopping-of-omni-7-until-3-31/
}.
The system's greatest strength is that consumers can receive goods at any of the approximately 20,000 Seven-Eleven stores throughout Japan.
This is comparable to the number of post offices (in operation) in Japan, such that, roughly speaking, anyone in an urban area can easily find a nearby Seven-Eleven store\footnote{
According to each site, the numbers of Seven-Eleven stores and post offices are 20,973 and 23,944, respectively, as of the end of June 2019.
}.
Many consumers who use the omnichannel choose to retrieve their goods at a nearby store rather than waiting for home delivery.
In most cases, these consumers would be able to walk from their homes to the store or stop in the store on their way home from work to pick up their parcels\footnote{See \citet{Terasawa1998} and \citet{Ishikawa_Nejo2002}.}.
\par
In our opinion, however, Omni 7 has not been entirely successful.
In fact, in October of 2016, Seven \& I Holdings Co. (the parent company of Seven-Eleven Japan) recognized significant impairments in the e-commerce field.
In February of 2018, the number of memberships in Omni 7 was about 7.85 million, and sales through the Omni 7 channel were approximately 108.7 billion yen---far from the target value of one trillion yen.
\par
Some consumer reviews mention the length of time between ordering an item and its in-store availability.
The most likely reason for this is that with BOPS, the existing ``magazine delivery system'' is used to ship to stores, while in other cases, the goods are delivered directly to consumers from fulfillment centers.
Convenience stores regularly order magazines, which are delivered to stores on a fixed schedule.
Stores use this delivery system for goods ordered through BOPS, and delivery usually takes several days. 
Using the established delivery system may reduce costs, but it may also mean that consumers have to wait for routine deliveries.
This latency is wasteful for consumers compared to the next-day delivery of EC vendors such as Amazon.com. 
\par
Consider the following hypothetical numerical example.
Suppose it takes up to seven days to deliver goods to a store through an existing magazine delivery system if the product is out of stock at the store.
On the other hand, in the case of postal delivery to a home, the goods are delivered the same day, but the fee is uniformly 400 yen\footnote{As a reference, the shipping rate charged by Amazon.co.jp as of July 2019 is 400 yen for orders worth less than 2,000 yen (except for remote islands). \\
https://www.amazon.co.jp/gp/help/customer/display.html?ie=UTF8\&nodeId=\\
642982\&ref\_=footer\_shiprates
}.
For simplification, suppose that the time cost for consumers who are waiting for a home delivery is equivalent to 100 yen per day.
Consumers view a retailer's site, confirm that BOPS is available for a product, and know that the product can be picked up at a nearby store. Then, they predict how many days it will take for the product to be received by the store.
If consumers expect that BOPS will take more than four days, they may opt not to use BOPS and instead go directly to the nearest store to purchase the item. In the case of a store stockout, they may elect to order the item online and receive the shipment the same day at a cost of 400 yen.
On the other hand, consumers who live far away from a store may decide to buy products directly from other internet vendors, as the cost of traveling to the store may be higher than 400 yen\footnote{
As a reference, in Japan, taxi fares are around 400--700 yen for the first two kilometers of a trip.
}.
Hence, for consumers to use BOPS, delivery to the store must be within four days and this may rely on special store deliveries that are costlier to the store than the existing magazine delivery system.
In this case, the retailers should not offer BOPS and instead induce consumers to visit the store directly to purchase items if the costs of the special deliveries may exceed the benefits.
\section{Model}
\subsection{Consumers Choices}
Consider a retailer who sells goods at a brick-and-mortar store and other EC venders such as Amazon.com.
The retailer then considers introducing a buy-online-and-pick-up-in-store (BOPS) system.
This option would enable consumers to order products on the website of the retailer and pick them up at the store.
For consumers, the advantage of BOPS would be that they have the certainty of receiving the good, while they face the risk of a stockout if they go directly to the store to purchase the item.
On the other hand, disadvantages of BOPS for consumers are that they must incur the cost of traveling to the designated store and they must wait for the product to arrive at the store if it is currently out of stock.
If this is the case, the retailer can immediately order more product for the store.
This scenario is contrary to \citet{Gao_Su2016}, in which BOPS is feasible \emph{only if} the good is in stock at the store.
Instead of using BOPS or going directly to the physical store to purchase the item, consumers can also buy the same item from another online company that charges a specified delivery fee.
We assume that the price of the good, $p$, is identical and exogenous for the two channels.
\par
Consequently, consumers can choose one of the following options:
(1) buy online directly,
(2) use BOPS and then go to the store to pick up the item,
or
(3) visit the store without using BOPS.
We suppose that the evaluation of the good, $v$, is identical for every consumer and that any consumer can buy the good, with $v - p$ assumed to be positive and sufficiently large throughout the use of this model.
\par
Here, to distinguish our model from \citet{Gao_Su2016}, we again emphasize that BOPS would not provide any information about the store's inventory. In the previous study, it was assumed that using BOPS for a product was only feasible when the product was in stock at the store, but our model does not make this assumption.
Therefore, some consumers may discover that a good is not currently available at the store, even after they see (on the retailer's website) that BOPS is available for that item. 
After consumers choose the BOPS option, they are notified immediately if the item is in stock at the store. Otherwise, they are informed they will need to wait several days for the item's arrival.
Consumers do not have the option of cancelling BOPS\footnote{
In Omni 7, cancellation is not possible after 30 minutes from the order confirmation}.
Furthermore, with BOPS, consumers must incur the cost of traveling to the store.
Hence, with BOPS, consumer utility is given by $v - p - t - \mu$, where $t$ is the travel cost and $\mu$ is disutility, or waiting time for the arrival of the good.
The disutility can be converted to a monetary amount, and for simplicity, we assume it can take two values: $\mu \in \{ 0, \bar{\mu} \}$.
It is zero when there is inventory, and $\bar{\mu} \geq 0$ when there is not inventory\footnote{
This binomial-variable formulation might seem overly simple.
One could easily extend it to multi- or continuous-variable models.
}.
Suppose that consumers have a belief $\hat{\xi}$ that the item is in stock.
Then, the expected utility for a consumer using the BOPS option is as follows:
\begin{equation}
    u_{b} = v - p - t - (1 - \hat{\xi})\bar{\mu},
\end{equation}
and the utility for the consumer buying directly online is the following:
\begin{equation}
    u_{o} = v - p - c_{o},
\end{equation}
where $c_{o}$ is the delivery fee for items purchased online.
In this model, we normalize the disutility (i.e., waiting time) in online shopping to zero.
\par
Consumers visiting the store without using BOPS would buy the good at the store if it is in stock. In a stockout, they would have to choose whether to buy it online or use BOPS.
The expected utility for the consumers in this scenario is as follows:
\begin{equation}
\label{eq:us}
    u_{s} = -t + \hat{\xi}(v - p) + (1 - \hat{\xi})\max\{ u_{o}, u_{b}^{0} \},
\end{equation}
where $u_{o}^{0}$ is the conditional expected utility under the condition of stockout:
\begin{equation}
\label{eq:ub0}
	u_{b}^{0} = v - p - t - \bar{\mu}.
\end{equation}
This formulation shows that consumers confirm a good is in stock and that they must wait with disutility $\bar{\mu}$.
\par
The sequence of consumer actions is as follows:
First, consumers see the BOPS option on the retailer's website.
The distance to a designated store determines the travel cost $t$; thus, each consumer knows his or her own value of $t$.
Before using the BOPS option, consumers receive information about the maximum waiting time, $\bar{\mu}$.
At the same time, they observe the delivery fee $c_{o}$ on the website of an online retailer such as Amazon.com\footnote{
These costs, $t$, $c_{o}$, and $\bar{\mu}$ correspond to $h_{s}$, $h_{o}$, and $h_{b}$ in \citet{Gao_Su2016}, respectively.
Notably, one of the contributions of this study is developing the concept for these cost parameters.
}.
Under the information for triplet $(t, c_{o}, \bar{\mu})$, consumers establish the belief $\hat{\xi}$ that an item is in stock at a store, and they determine whether to use BOPS, not to use BOPS and go directly to the store to purchase the item, or to buy the item online directly.
Note that with the BOPS option, consumers receive information about the waiting time, $\mu$, \emph{only after} they choose to use BOPS.
\par
Therefore, we arrive at the following proposition\footnote{The proofs for propositions are shown in Appendix.}:
\begin{prop}
	(1) Given $\hat{\xi}$, consumers proceed with choices in the following manner:
Case I ($\hat{\xi} = 1$).
If $0 \leq \bar{\mu} \leq c_{o}$, consumers with $0 \leq t \leq c_{o}$ use BOPS, and others buy online directly.
If $c_{o} < \bar{\mu}$, consumers with $0 \leq t \leq c_{o}$ visit the store without using BOPS, and others buy online directly.
Case II ($0 < \hat{\xi} < 1$).
If $0 \leq \bar{\mu} \leq c_{o}$, consumers with $0 \leq t \leq c_{o} - (1 - \hat{\xi})\bar{\mu}$ use BOPS, and others buy online directly.
If $\bar{\mu} > c_{o}$, consumers with $0 \leq t \leq \hat{\xi}c_{o}$ do not use BOPS and go directly to the store, and others buy online directly.
Case III ($\hat{\xi} = 0$).
If $0 \leq \bar{\mu} \leq c_{o}$, consumers with $0 \leq t \leq c_{o} - \bar{\mu}$ use BOPS, and others buy online directly.
If $c_{o} < \bar{\mu}$, all consumers buy online directly.
\
(2) If consumers not using BOPS cannot find a good, they buy it online afterwards.
\end{prop}
\par
Figure 1 shows the optimal consumer choices.
\par
\vspace{\baselineskip}
\begin{figure}[h]
\centering
\includegraphics[width=9.3cm,bb=0 0 1324 540]{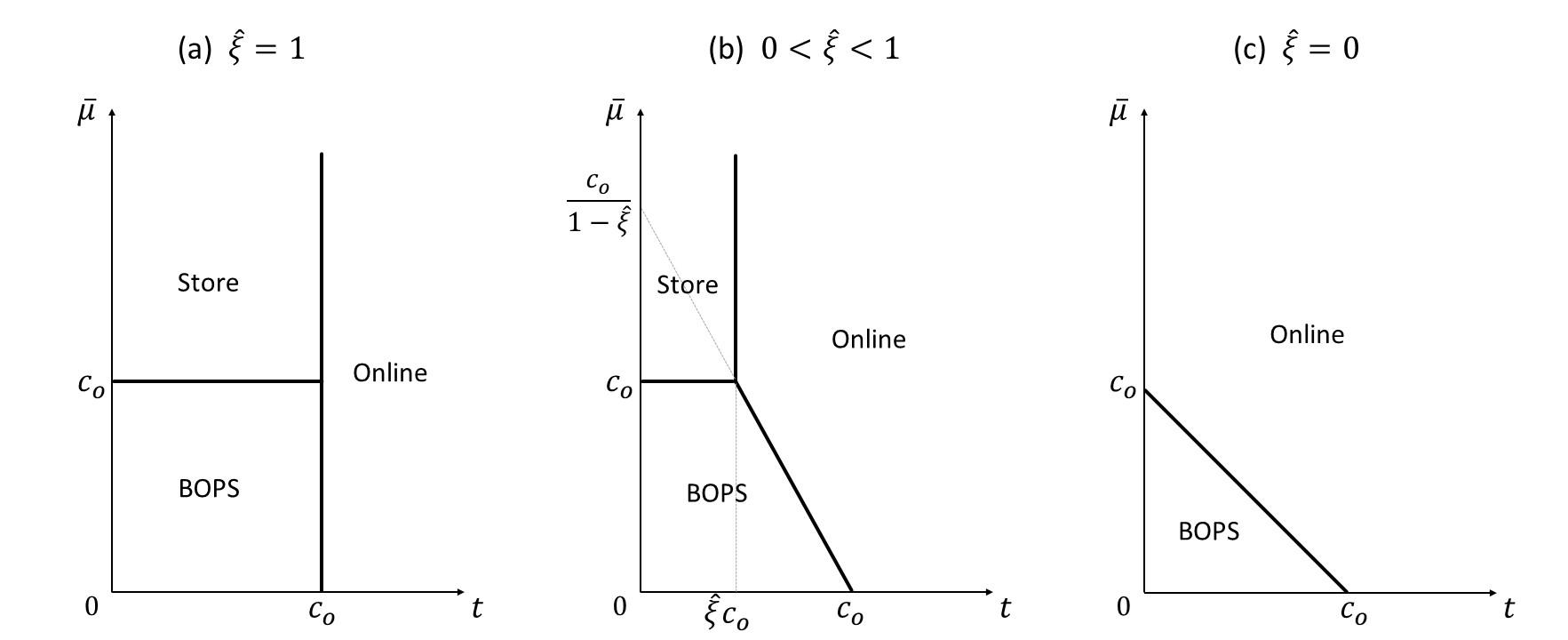}
\caption{Consumers Choice with BOPS}
\end{figure}
\vspace{\baselineskip}
\noindent
Figure 1(a), (b), and (c) correspond to Case I, II, and III in Proposition 1.
In the figures, ``BOP'' regions correspond to consumers who use BOPS,``Store'' to consumers who go directly to the store to purchase items without using BOPS, and ``Online'' to consumers who buy items directly online.
\par
Proposition 1(1)has two main implications. 
First, in the region with $\bar{\mu} > c_{o}$, for any $\hat{\xi}$ and $t$, BOPS is not useful.
Second, for any $\hat{\xi}$ and $\bar{\mu}$, consumers with $t > c_{o}$ buy online directly.
Thus, if the disutility or waiting time for BOPS is sufficiently large compared to the online delivery fee, then BOPS is not useful, regardless of inventory probability.
This implication highlights one of the main reasons that some omnichannel initiatives are not successful\footnote{
One can find online customer reviews of Omni 7 that complain about having to wait too long for the arrival of their goods.
}.
Next, we find that the retailer can acquire consumers at most $t = c_{o}$.
BOPS is useless for a retailer if the cost of raising inventory probability, $\hat{\xi}$, is zero, because the retailer can acquire $c_{o}$ consumers by inducing them to the store without reducing $\bar{\mu}$.
\par
Furthermore, if consumers visit the store to purchase items without using BOPS, they find $u_{o} > u_{b}^{0}$.
As shown in Proposition 1(2), if an item is out of stock, consumers buy it online but do not use BOPS.
\par
To illustrate the number of consumers a retailer can acquire by offering a BOPS option, in Figure 2, we show the consumer choices before BOPS was introduced. 
\par
\vspace{\baselineskip}
\begin{figure}[h]
\centering
\includegraphics[width=9.3cm,bb=0 0 1324 540]{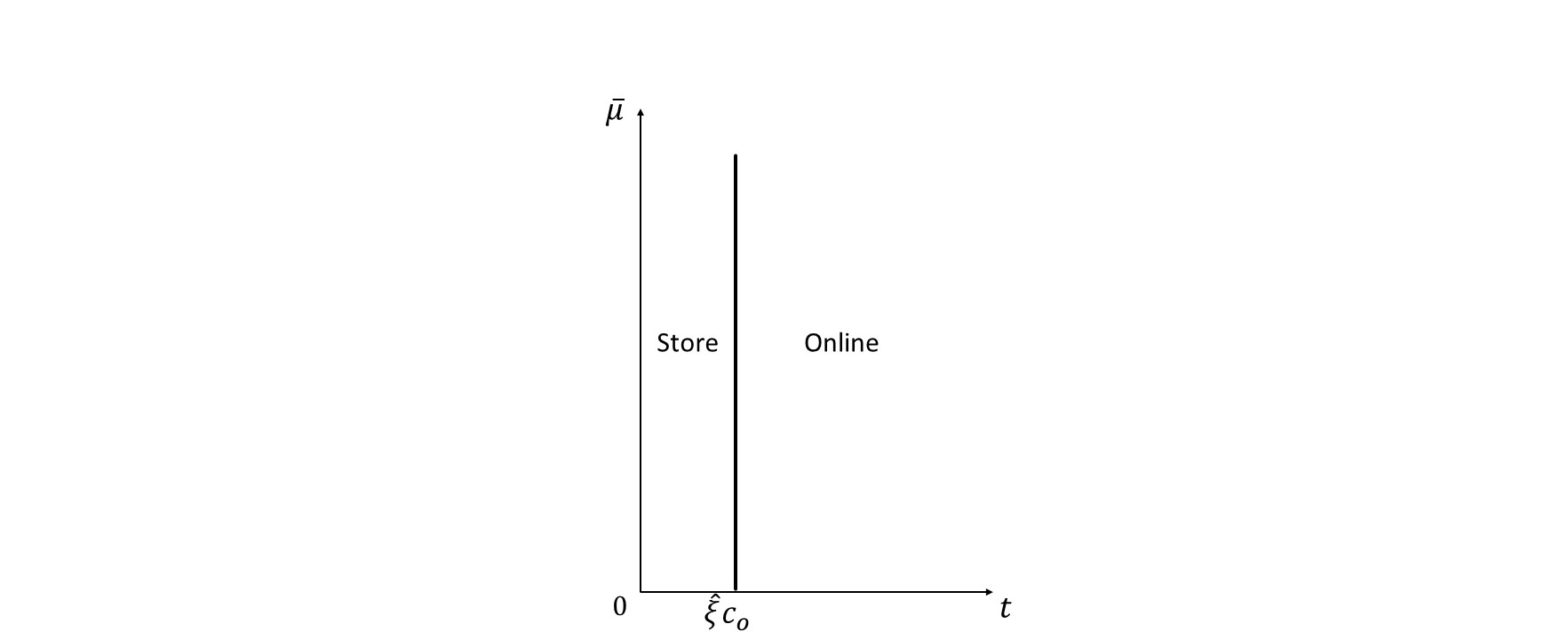}
\caption{Consumers Choice without BOPS}
\end{figure}
\vspace{\baselineskip}
\noindent
Figure 2 shows that for any $\bar{\mu}$, consumers with $0 \leq t \leq \hat{\xi}c_{o}$ visit the store to purchase the item, and others buy it online directly.
Compared to Figure 1, we do not find any new consumers by introducing BOPS if $\hat{\xi} = 1$.
The retailer can acquire new consumers equivalent to $(1 - \hat{\xi})c_{o}^{2}/2$ with $0 < \hat{\xi} < 1$ and $c_{o}^{2}/2$ with $\hat{\xi} = 0$.
We show without proof that BOPS can attract more new consumers with a smaller $\hat{\xi}$, as in the following:
\begin{prop}
  For $0 \leq \hat{\xi} \leq 1$, the retailer acquires $(1 - \hat{\xi})c_{o}^{2}/2$ new consumers by introducing BOPS.
\end{prop}
In other words, BOPS can deter more consumers from online shopping when their expectation is that a good has a low probability of being in stock.
Even if a good is out of stock, thus $\hat{\xi} = 0$, the retailer can successfully induce some consumers to use BOPS and thereby, visit the store.
For these consumers, the cost of traveling to the store plus the disutility (when using BOPS) is less than the online delivery fee.
\subsection{Optimal Stock Probabilities and Equilibrium}
In this section, we have optimal inventory probabilities and equilibria.
For the equilibrium concept in this model, we adopt rational expectations (RE) equilibrium following \citet{Gao_Su2016}.
Roughly speaking, RE equilibrium in this model is such that any consumer chooses optimal actions based on beliefs about inventory levels, consistent with the retailer's optimal inventory strategies\footnote{A rigorous discussion of RE equilibrium can be found in \citet{Stokey1981}.}.
\par
Suppose that consumers are located in space with travel cost $[0, \infty)$ and a density of one, and they each buy one unit of a good.
From Proposition 1, the demand of the good, $0 \leq \hat{\xi} \leq 1$, is as follows\footnote{
Contrary to \citet{Gao_Su2016}, demand is not a random variable here.
Their model derived the optimal inventory as the inverse function of the distribution of the demand.
}:
\begin{equation}
\label{eq:Demand}
  D =
  \begin{cases}
    c_{o} - (1 - \hat{\xi})\bar{\mu} & (0 \leq \bar{\mu} \leq c_{o}) \\
    \hat{\xi}c_{o} & (c_{o} < \bar{\mu})
  \end{cases}
\end{equation}
Given the demand, the retailer maximizes profit with respect to two variables: inventory level $q$ and maximum disutility $\bar{\mu}$.
We let $\pi(q, \bar{\mu})$ denote the profit function.
\par
Here, we define RE equilibrium as follows:
\begin{defn}
 RE equilibrium is a set, $(q, \bar{\mu}, \hat{\xi})$, satisfying the following conditions:
  \begin{enumerate}
    \item Given $\hat{\xi}$, any consumer behaves as shown in Proposition 1,
    \item Given the consumer behaviors in Proposition 1, $(q, \bar{\mu}) = \mathrm{argmax}_{(q', \bar{\mu}')}\pi(q', \bar{\mu}')$,
    \item $\hat{\xi} = \min\{ q/D, 1 \}$, where $D$ is defined as in Eq.(\ref{eq:Demand}).
  \end{enumerate}
\end{defn}
We let $\xi(q, \bar{\mu})$ denote the inventory probability that satisfies the third condition of the above definition as a function of $q$ and $\bar{\mu}$.
The inventory probability function is solved as follows:
\begin{equation}
\label{eq:xi}
  \xi(q, \bar{\mu}) =
  \begin{cases}
    \min\left\{
      \frac{\displaystyle -(c_{o} - \bar{\mu}) + \sqrt{(c_{o} - \bar{\mu})^{2} + 4\bar{\mu}q}}{\displaystyle 2\bar{\mu}},
      \ 1
    \right\} & (0 \leq \bar{\mu} \leq c_{o}) \\
    \min\left\{
      \sqrt{\frac{\displaystyle q}{\displaystyle c_{o}}},
      \ 1
    \right\} & (c_{o} < \bar{\mu})
  \end{cases}
\end{equation}
The profit function can be denoted as follows:
\begin{equation}
  \pi(q, \bar{\mu}) =
  \begin{cases}
    p\Bigl\{
      c_{o} -
      \bigl[
        1 - \xi(q, \bar{\mu})
      \bigr]\bar{\mu}
    \Bigr\}
    - cq - C(\bar{\mu}) & (0 \leq \bar{\mu} \leq c_{o}) \\
    p\xi(q, \bar{\mu})c_{o}
    - cq - C(\bar{\mu}) & (c_{o} < \bar{\mu}) \\
  \end{cases}
\end{equation}
\par
Here, we identify the cost of reducing $\bar{\mu}$, i.e., $C(\bar{\mu})$.
Clearly, $C(\bar{\mu}_{1}) < C(\bar{\mu}_{2})$ for any $\bar{\mu}_{1}$, $\bar{\mu}_{2}$ such that $\bar{\mu}_{1} > \bar{\mu}_{2}$.
In the following section, we simply suppose $C(\bar{\mu}) \equiv k(M - \bar{\mu})$.
Thus, $k$ is the marginal cost for reducing $\bar{\mu}$ by one unit or one day, and $M$ is a fixed number.
Hereafter, we assume $0 \leq k < p$.
If the retailer does not wish to reduce $\bar{\mu}$ at all, $\bar{\mu}$ must be $M$, and we assume $M > c_{o}$.
Thus, consumers must incur a disutility greater than the online delivery cost if the retailer makes no effort.
\par
Finally, we can maximize the retailer's profit.
It is clear from Proposition 1 and Figure 1 that, for a given $\bar{\mu}$, any consumer buying from the retailer uses BOPS if $0 \leq \bar{\mu} \leq c_{o}$ and visits the store to purchase items without BOPS if $c_{o} < \bar{\mu} \leq M$.
The following proposition shows the optimal inventory level to maximize the profit, $\pi(q, \bar{\mu})$.
\begin{prop}
  For $\bar{\mu}$, the retailer's optimal solutions with respect to $q$ and optimal profits are shown in Table 1.
\end{prop}
In this table, the upper and lower parts represent $p \geq 2c$ and $c < p < 2c$, respectively.
BOPS I, II, and III correspond to solutions when consumers use BOPS, and Stores I and II to solutions when consumers do not use BOPS.
In BOPS I, III, and the Store I regions, there are corner solutions: the optimal inventory level is $q = D$ in BOPS I and Store I, but in BOPS III, it is $q = 0$.
When consumers do not use BOPS, $q = 0$ is not optimal; thus, there does not exist a ``Store III ($\xi = 0$)'' region.
Clearly, the retailer cannot sell goods in the store when the inventory probability is zero.
Figure 3 shows the solutions in space $(\bar{\mu}, p)$.
\par
\vspace{\baselineskip}
\begin{figure}[h]
\centering
\includegraphics[width=9.3cm,bb=0 0 960 540]{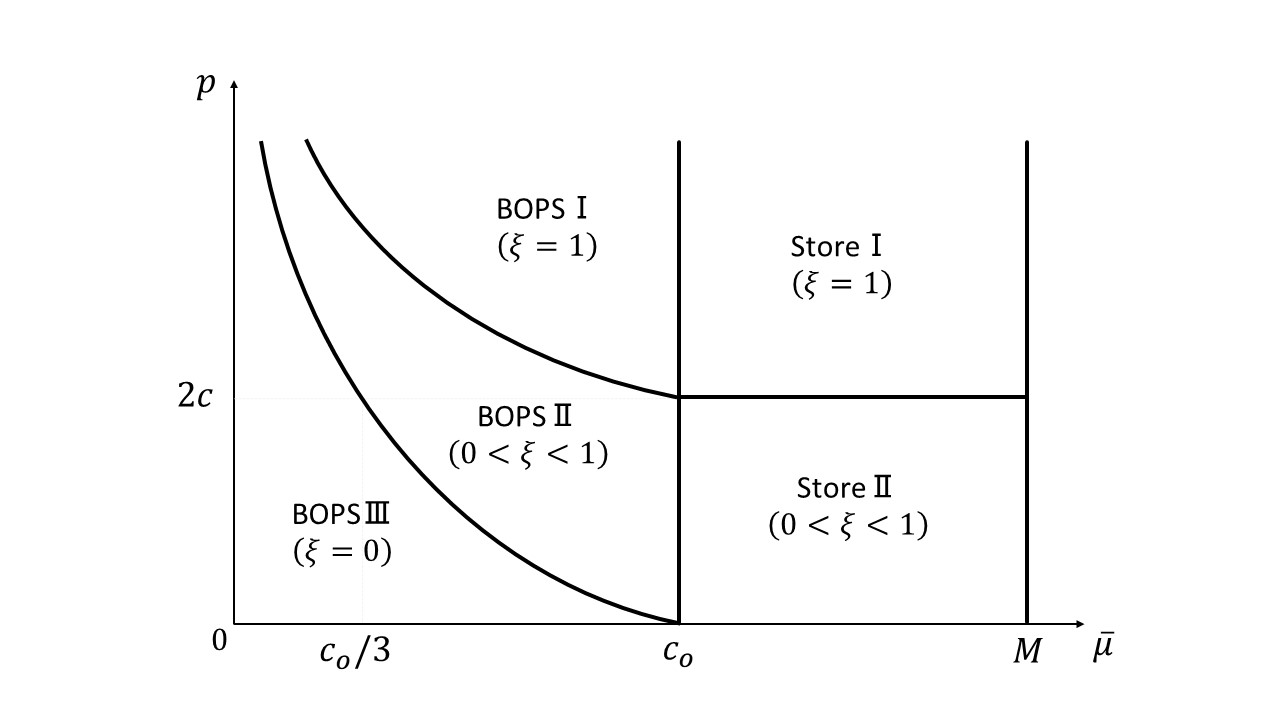}
\caption{Inventory Solution Regions}
\end{figure}
\vspace{\baselineskip}
\par
Now, we are ready to find the optimal disutility $\bar{\mu}$ in each of the regions in Figure 3.
The following proposition identifies the optimal solutions of $\bar{\mu}$ in the regions, say, \emph{local solutions}.
\begin{prop}
 The optimal solutions of $\bar{\mu}$ are as follows: $\bar{\mu} = c_{o}$ in BOPS I, $\bar{\mu} = (c/(p - c))c_{o}$ when $p \geq 2c$ and $\bar{\mu} = c_{o}$ when $c < p < 2c$ in BOPS II, $\bar{\mu} = 0$ in BOPS III, and $\bar{\mu} = M$ in Store I and Store II.
\end{prop}
Regions of the local solutions in Proposition 4 are shown in Figure 4.
\par
\vspace{\baselineskip}
\begin{figure}[h]
\centering
\includegraphics[width=9.3cm,bb=0 0 960 540]{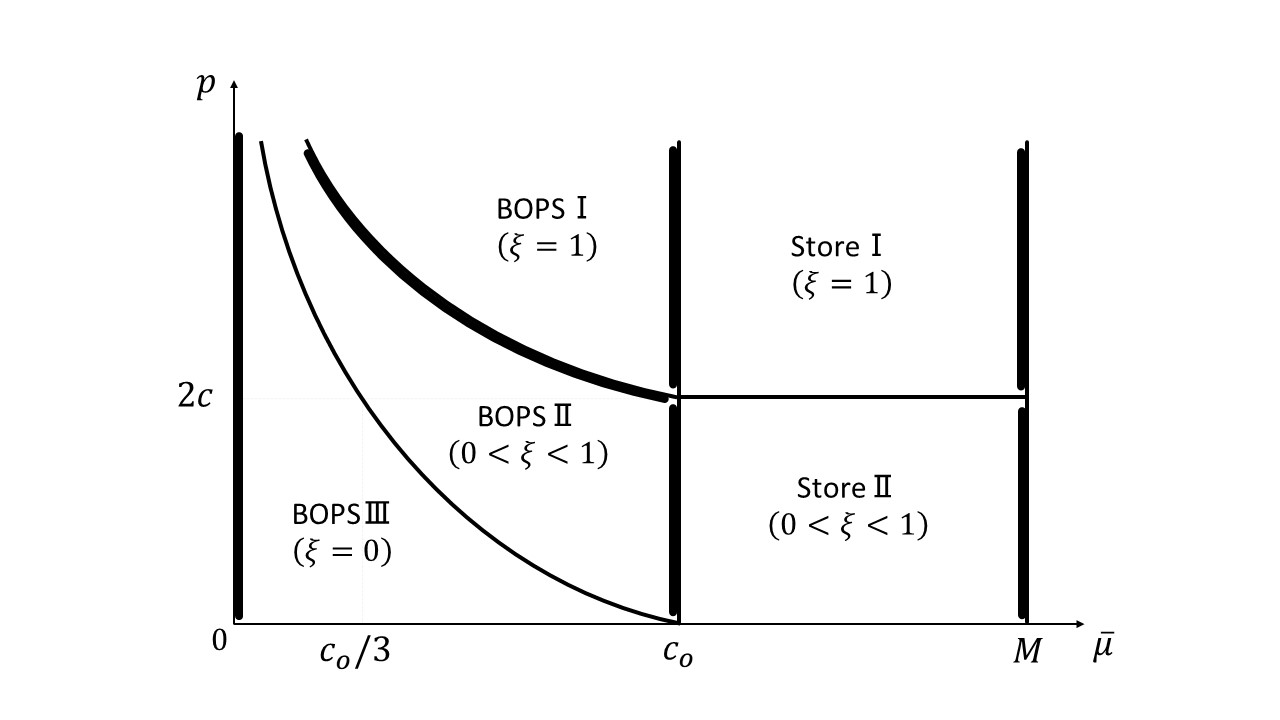}
\caption{Optimal $\bar{\mu}$ in Solution Regions}
\end{figure}
\vspace{\baselineskip}
\noindent
It should be noted that there is no interior solution for $\bar{\mu}$ in any region.
Furthermore, $\bar{\mu}$ must be $0$, $c_{o}$, and $M$, except for in BOPS II.
\par
Finally, we have \emph{global solutions} of $\bar{\mu}$ as follows:
\begin{prop}
When $p \geq 2c$, the optimal solution is as follows: $\bar{\mu} = 0$ (BOPS III) if $k \leq (c_{o}/M)c$ and $\bar{\mu} = M$ (Store I) if $k > (c_{o}/M)c$. When $c < p < 2c$, the optimal solution is as follows: $\bar{\mu} = 0$ (BOPS III) if $k \leq (1 - p/4c)pc_{o}/M$ and $\bar{\mu} = M$ (Store II) if $k > (1 - p/4c)pc_{o}/M$.
\end{prop}
In short, we can summarize the optimal $\bar{\mu}$ as $0$ or $M$, and $\bar{\mu} = 0 $ if $q = 0$.
\par
The equilibrium sets of $(q, \bar{\mu})$ are shown in space $(c, k)$ as shown in Figure 5.
\par
\vspace{\baselineskip}
\begin{figure}[h]
\centering
\includegraphics[width=9.3cm,bb=0 0 960 540]{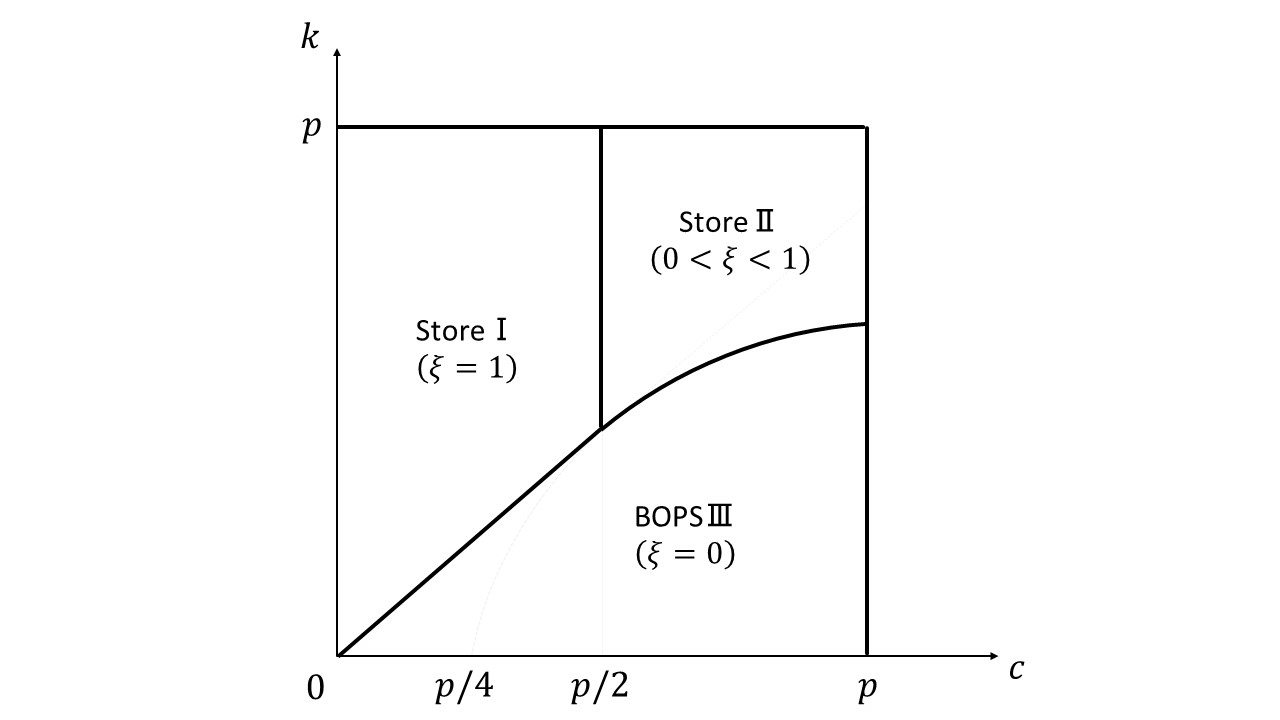}
\caption{Equilibrium Regions}
\end{figure}
\vspace{\baselineskip}
\noindent
In this figure, $(q, \bar{\mu}) = (0, 0)$ in BOPS III, $(q, \bar{\mu}) = (c_{o}, M)$ in Store I, and $(q, \bar{\mu}) = ((p/2c)^{2}c_{o}, M)$ in equilibrium.
\section{Discussion}
The above results have some implications.
\par
First, the model has two types of equilibrium: all consumers use BOPS and all consumers do not use BOPS.
Then, the retailer would offer a BOPS option only when consumers are willing to use it, and the decision depends on two types of costs, $c$ and $k$.
When $k$ is greater than $c$, it is not profitable for retailers to induce consumers to use BOPS by reducing the BOPS waiting times.
In this case, the retailer does not offer the BOPS option and instead, induces consumers to visit the store, which cancels the retailer's incentive to reduce the BOPS waiting time, $\bar{\mu} = M$. On the other hand, the retailer wishes to increase the inventory probability, $\xi$, by increasing the inventory level, $q$.
When $k < c$, increasing the inventory level does not benefit the retailer.
Then, the retailer sets the inventory probability to zero and reduces the waiting time to induce the consumers to use BOPS, $\bar{\mu} = 0$.
\par
Second, in the Store region equilibrium, there may be an interior or corner solution for $q$, but in the BOPS region equilibrium, there is only a corner solution.
There is no interior BOPS equilibrium in this model because of the assumption about the cost of reducing the BOPS waiting time:
We assume $C(\bar{\mu}) = k(M - \bar{\mu})$ and $k < p$, which results in $\bar{\mu} = 0$ as the unique solution.
Then, the retailer has to incur the cost, $kM$, avoiding the cost, $cq$, which means $q = 0$.
Note that these results depend on the cost formulation assumption, and we can derive other results with interior BOPS solutions when assuming another formulation.
\par
Finally, these results suggest a strong motivation to achieve a high density of stores so that the retailer can fulfill replenishments at a lower cost.
We think this is the motivation for the ``dominant system'' established by Seven-Eleven Japan\footnote{
According to \citet{Ishikawa_Nejo2002}, the dominant system entails ``focusing operations in certain areas which are then crowded with multiple stores'' (p.51).
}.
In conclusion, we stress in this discussion that a successful omnichannel initiative must be tied to an efficient fulfillment system.
\section{Numerical Example}
We now examine the above results in numerical examples.
Suppose $\bar{\mu} = \{ 0, 1, 2, \ldots, 6, 7 \}$, $c_{o} = 4$, and $M = 7$.
In other words, consumers must incur a delivery fee equivalent to four days of waiting, and they have to wait for seven days if the retailer does not reduce the BOPS waiting time.
The retailer can reduce the waiting time to zero.
\par
We calculate equilibrium solutions in Store and BOPS regions for four cases, as shown in Figure 6.
\par
\vspace{\baselineskip}
\begin{figure}[h]
\centering
\includegraphics[width=9.7cm,bb=0 0 960 666]{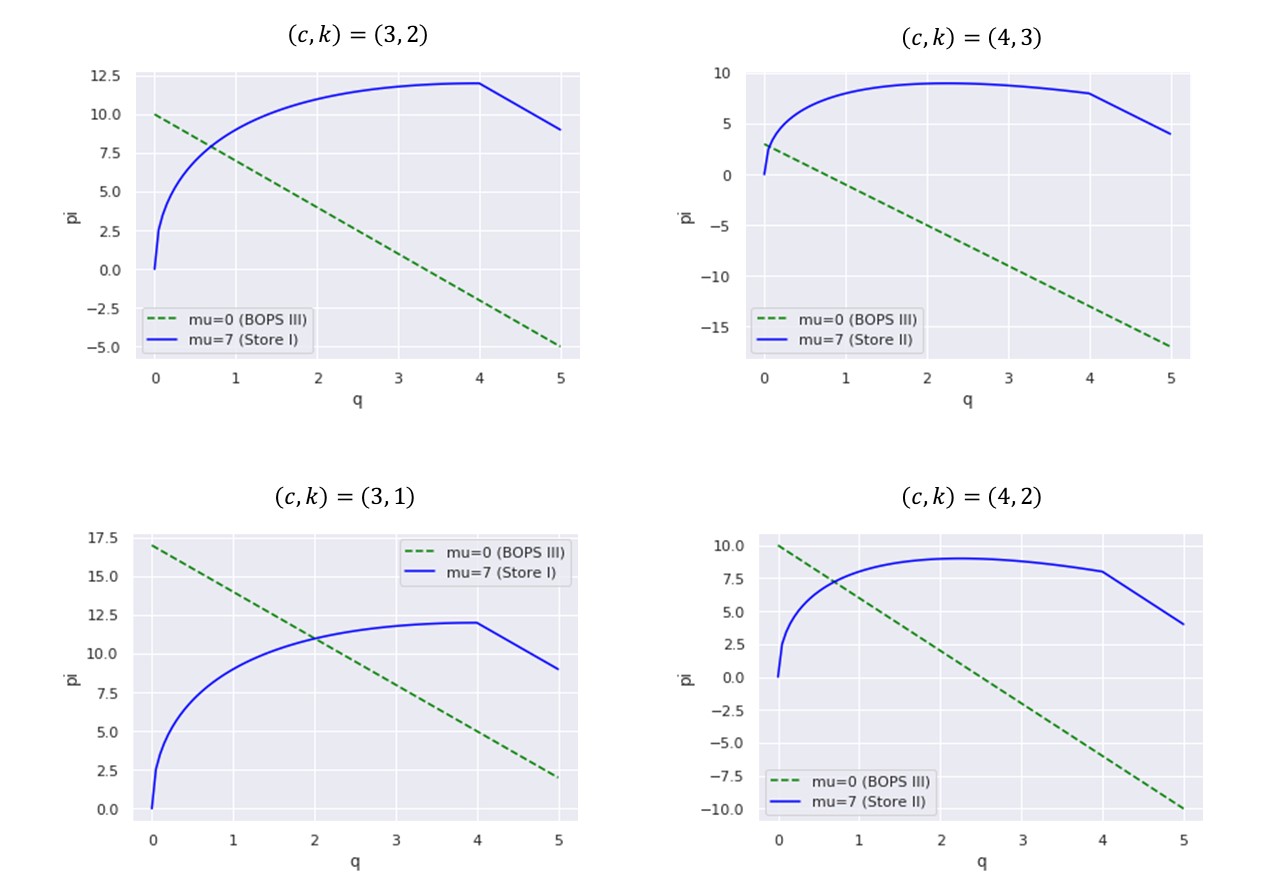}
\caption{Numerical Examples}
\end{figure}
\vspace{\baselineskip}
\noindent
In the figure, the upper left and upper right diagrams show equilibrium with corner solutions (Store I) and interior solutions (Store II) in Store region, respectively.
In these cases, $k$ is significant for the retailer compared to $c$, and BOPS is not useful.
Furthermore, when $c$ is significant, the optimal solution for $q$ cannot be one, which yields the interior solution.
In any case, the BOPS option is not profitable, and the retailer does not offer it to consumers.
\par
The lower diagrams show equilibrium in BOPS region (BOPS III).
In this case, $k$ is not significant compared to $c$, and the retailer chooses $q = 0$.
The consumers rationally expect that the good is \emph{out of stock} in the store, and they must use BOPS before visiting the store.
In a stockout, the store is merely a ``pickup location'' for the item.
Consequently, we can interpret \emph{pickup lockers} as a solution for the new fulfillment initiative in the omnichannel era\footnote{
Amazon.com offers \emph{Amazon Hub Locker}, and Walmart offers \emph{Walmart Pickup Tower}.
}.
\section{A Policy Suggestion}
Recently, within the Japanese convenience store industry, a controversy erupted regarding around-the-clock services, which was triggered by a dispute between Seven-Eleven Japan and one of its franchisees.
In February of 2019, a franchise store in Osaka decided to close between 1 and 6 in the morning because of the labor shortage\footnote{
``Japan's 24-hour convenience stores struggle to keep doors open all night due to labor crunch,'' \emph{The Japan Times}, February 27, 2019.
}.
\par
In their entire history spanning over 40 years, Japanese convenience stores---including Seven-Eleven Japan---have offered 24-hour services.
However, due to the recent labor shortage in Japan, it has become challenging to keep stores open at night, presenting a burden for store owners.
Although Seven-Eleven Japan agreed to shorten operating hours for stores in July of 2019, it seems not to have changed the 24-hour operation policy overall, according to company officials\footnote{
``Seven-Eleven Japan offers to allow store to end 24-hour operations,'' \emph{The Japan Times}, July 12, 2019.
}.
\par
In the context of this study, the operating hours of convenience stores is a critical issue for BOPS feasibility.
Convenience stores in Japan are open all day, while the Japan Post Office usually only operates from 9 a.m. to 7 p.m. (weekdays), meaning that BOPS has an advantage over online shopping, which depends on the postal service.
Shortening the operating hours of convenience stores could significantly impair the superior position of BOPS or Omni 7, creating a strong motivation to maintain 24-hour operations and thus, maximize BOPS convenience.
\par
In the ``convenience-store-opening-hours controversy,'' it seems that this point of view is lacking.
The 24-hour operating schedule enables consumers to pick up goods at any store, anytime.
Besides, the logistics of meeting demand on time may require store operation even at night.
We stress that the 24-hour operations problem should be discussed, considering both the necessity of securing a workforce for sales and the benefits of new systems, including omnichannels.
\section{A Simple Extension}
In this section, consider a simple, dynamic, cross-channel fulfillment model for a future extension.
We do not suppose the convenience scenario, and each store can replenish the inventories by using other stores, which have extra capacity.
\citet{Harsha_Subramanian_Uichanco2019} study this kind of model, in which a store's inventory can be utilized for fulfillment at other stores when the e-commerce fulfillment center is out of stock.
Here, for the sake of simplicity, let us consider a model in which there are two stores, 0 and 1. If the good in one store is not sold out in the previous period, it can be used for BOPS fulfillment in the other store, which reduces $k$ in the above model.
\par
Every week, each store decides to adopt BOPS III or not.
At the beginning of each week, Store 0 (Store 1) observes the inventory status of Store 1 (Store 0), and it chooses BOPS III ($\bar{\mu} = 0$, $q = 0$) if the good is in stock, or Store I ($\bar{\mu} = M$, $q = c_{o}$) otherwise.
Then, the weekly demand of the store that chooses Store I, $D$, arises. If $D < q$, with probability $r$, the unsold goods can be used in the other store in the next period.
This process generates a sequence of binomial states for each store.
\par
Table 2 shows the simulation result for 20 weeks of three probability cases in this scenario.
In this table, $S$ and $B$ represent Store I and BOPS III, respectively.
It is not surprising that BOPS III arises more often at greater $r$.
\par
This dynamic scenario can be extended to more complex fulfillment systems, such as those in \citet{Harsha_Subramanian_Uichanco2019}.
For example, we can suppose a model in which each store is randomly located within random distances of one another.
In this case, there arise several levels of $k$, because the nearest store's location is different for each store.
\par
This extension is open to future studies.
\section{Conclusion}
In this paper, we study BOPS following \citet{Gao_Su2016} to develop another stylized model.
We focus on consumer choices regarding travel and delivery costs and waiting time, and we show that there exist two types of equilibrium: using BOPS and not using BOPS, even without the information effect imposed on \citet{Gao_Su2016}.
We restrict the model to one in which a retailer sells goods only through a physical store. Thus, we did not have to consider the cross-selling effect.
\par
The above model can be extended by relaxing the assumptions in the model.
First, one could develop a model in which the retailer operates both physical and online stores.
Selling in stores has the advantage that consumers might make additional purchases.
The cross-selling effect should be examined in the two-channel model.
Second, we could use the ``unit'' model, with one retailer as a multi-channel and dynamic-fulfillment model, as in \citet{Harsha_Subramanian_Uichanco2019} to depict more realistic omnichannel strategies.
The cross-channel fulfillment model could be established, incorporating one channel as in this model.
Finally, price strategies are not addressed in this paper and need to be considered.
The retailer could charge varying prices in different areas or periods.
With additional fulfillment and price strategies in future studies, we would have richer formulations to more deeply understand the dynamic realm of omnichannels.
\appendix
\section{Proof of Proposition 1}
Here, we only derive the results in Proposition 1 for Case II ($0 < \bar{\mu} < 1$).
To determine consumer choices, we compare the utilities in the manner of backward induction.
In the first stage, we obtain $\max\{ u_{o}, u_{b}^{0} \}$ in Eq.(\ref{eq:us}) to determine $u_{s}$ and compare $u_{o}$, $u_{b}$, and $u_{s}$ in the second stage.
(1) If $t \geq c_{o} - \bar{\mu}$, $u_{o} = \max\{ u_{o}, u_{b}^{0} \}$ and $u_{s} = v - p - t - (1 - \hat{\xi})c_{o}$.
Then, if $0 \leq \bar{\mu} \leq c_{o}$, $u_{b} \geq u_{s}$, and vice versa.
When $0 \leq \bar{\mu} \leq c_{o}$, $u_{b} \geq u_{o}$ if $0 \leq t \leq c_{o} - (1 - \hat{\xi})\bar{\mu}$, and $u_{o} > u_{b}$ otherwise.
When $\bar{\mu} > c_{o}$, $u_{s} \geq u_{o}$ if $0 \leq t \leq \hat{\xi}c_{o}$, and $u_{o} > u_{s}$ otherwise.
(2) If $t < c_{o} - \bar{\mu}$, $u_{b}^{0} = \max\{ u_{o}, u_{b}^{0} \}$ and $u_{s} = v - p - (2 - \hat{\xi})t - (1 - \hat{\xi})\bar{\mu}$.
In this case, we just have to consider $u_{b} \geq u_{s}$ because $t$ cannot be negative.
For any $t \in [0, c_{o} - \bar{\mu}]$, $u_{b}$ cannot be smaller than $u_{o}$.
Finally, we combine (1) and (2) to obtain Proposition 1, Case II.
For Cases I and III, we can proceed in the same manner.
$\blacksquare$
\section{Proof of Proposition 3}
We only show the BOPS result in Table 1 (i.e., $0 \leq \bar{\mu} \leq c_{o}$).
The inventory probability function is shown in the upper part in Eq. (\ref{eq:xi}).
\par
First, suppose $0 < \xi < 1$.
Maximizing $\pi(q, \bar{\mu})$ with respect to $q$, we obtain the optimal solution, $q = [(p\bar{\mu}/c)^{2} - (c_{o} - \bar{\mu})^{2}]/4\bar{\mu}$ and $\pi = (c/4\bar{\mu})\{[(p - c)/c]\bar{\mu} + c_{o}\}^{2} - C(\bar{\mu})$.
Then, the optimal inventory probability is $\xi = \{[(p + c)/c]\bar{\mu} - c_{o}\}/2\bar{\mu}$.
The condition for $0 < \xi < 1$ is $[c/(p + c)]c_{o} < \bar{\mu} < \min\{ [c/(p - c)]c_{o}, c_{o} \}$.
If $p \geq 2c$, then $c_{o} \geq [c/(p - c)]c_{o}$, and vice versa.
Because $\xi = 1$ only if $\bar{\mu} \geq [c/(p - c)]c_{o}$, $\xi$ cannot be one when $p < 2c$.
\par
Next, note that $\xi = 1$ only if $q \geq c_{o}$.
Then, the solution is $q = c_{o}$, and the profit is $\pi = (p - c)c_{o} - C(\bar{\mu})$ in this case.
Finally, $\xi = 0$ only if $q = 0$, and the profit is $\pi = p(c_{o} - \bar{\mu}) - C(\bar{\mu})$.
$\blacksquare$
\section{Proof of Proposition 4}
It is obvious that the optimal solution of $\bar{\mu}$ is $M$ in Stores I and II.
Then, we only examine BOPS regions.
In BOPS I, the profit is $\pi = (p - c)c_{o} - k(M - \bar{\mu})$, and the optimal $\bar{\mu}$ is $c_{o}$, because $\bar{\mu}$ cannot be larger than $c_{o}$ in this region.
In BOPS III, the profit is $\pi = p(c_{o} - \bar{\mu}) - k(M - \bar{\mu})$.
The optimal solution is $\bar{\mu} = 0$ because of the assumption, $p > k$.
In BOPS II, there exists no interior solution for $\bar{\mu}$.
This is because the profit $\pi = (c/4\bar{\mu})\{[(p - c)/c]\bar{\mu} + c_{o}\}^{2} - C(\bar{\mu})$ is a convex function of $\bar{\mu}$.
In fact, $\pi_{\mu\mu} = 4c_{o}^{2}c/\bar{\mu}^{2} > 0$.
Let $\pi(\bar{\mu})$ denote this function.
The domain is $[c/(p + c)]c_{o} < \bar{\mu} < \min\{ [c/(p - c)]c_{o}, c_{o} \}$, and we can easily show that $\pi([c/(p - c)]c_{o}) > \pi([c/(p + c)]c_{o})$ when $p \geq 2c$ and $\pi(c_{o}) > \pi([c/(p + c)]c_{o})$ when $p < 2c$.
$\blacksquare$
\section{Proof of Proposition 5}
When $p \geq 2c$, the optimal profits in Store I, BOPS I, and BOPS II are $\pi = (p - c)c_{o}$, $\pi = (p - c)c_{o} - k(M - c_{o})$, and $\pi = (p - c)c_{o} - k\{ M - [c/(p - c)]c_{o} \}$, respectively.
Obviously, the profit in Store I dominates that of the others.
In this case, the optimal profit in BOPS III is $\pi = pc_{o} - kM$.
Then, the profit in BOPS III dominates that of Store I if $k \leq (c_{o}/M)c$, and vice versa.
Similarly, the profit in BOPS III is the maximum if $k \leq (1 - p/4c)pc_{o}/M$, and that of Store I is so otherwise.
$\blacksquare$
\par
\vspace{2\baselineskip}
\bibliographystyle{elsarticle-harv} 
\bibliography{kusuda_BOP_arXiv}

\begin{thebibliography}{17}
\expandafter\ifx\csname natexlab\endcsname\relax\def\natexlab#1{#1}\fi
\providecommand{\url}[1]{\texttt{#1}}
\providecommand{\href}[2]{#2}
\providecommand{\path}[1]{#1}
\providecommand{\DOIprefix}{doi:}
\providecommand{\ArXivprefix}{arXiv:}
\providecommand{\URLprefix}{URL: }
\providecommand{\Pubmedprefix}{pmid:}
\providecommand{\doi}[1]{\href{http://dx.doi.org/#1}{\path{#1}}}
\providecommand{\Pubmed}[1]{\href{pmid:#1}{\path{#1}}}
\providecommand{\bibinfo}[2]{#2}
\ifx\xfnm\relax \def\xfnm[#1]{\unskip,\space#1}\fi
%Type = Article
\bibitem[{Allon and Bassamboo(2011)}]{Allon_Bassamboo2011}
\bibinfo{author}{Allon, G.}, \bibinfo{author}{Bassamboo, A.},
  \bibinfo{year}{2011}.
\newblock \bibinfo{title}{Buying from the babbling retailer? {T}he impact of
  availability information on customer behavior}.
\newblock \bibinfo{journal}{Management Science} \bibinfo{volume}{57},
  \bibinfo{pages}{713--726}.
%Type = Article
\bibitem[{Bell et~al.(2014)Bell, Gallino and Moreno}]{Bell_Gallino_Moreno2014}
\bibinfo{author}{Bell, D.R.}, \bibinfo{author}{Gallino, S.},
  \bibinfo{author}{Moreno, A.}, \bibinfo{year}{2014}.
\newblock \bibinfo{title}{How to win in an omnichannel world}.
\newblock \bibinfo{journal}{MIT Sloan Management Review} \bibinfo{volume}{56},
  \bibinfo{pages}{45--53}.
%Type = Article
\bibitem[{Brynjolfsson et~al.(2013)Brynjolfsson, Hu and
  Rahman}]{Brynjolfsson_Hu_Rahman2013}
\bibinfo{author}{Brynjolfsson, E.}, \bibinfo{author}{Hu, Y.J.},
  \bibinfo{author}{Rahman, M.S.}, \bibinfo{year}{2013}.
\newblock \bibinfo{title}{Competing in the age of omnichannel retailing}.
\newblock \bibinfo{journal}{MIT Sloan Management Review} \bibinfo{volume}{54},
  \bibinfo{pages}{23--29}.
%Type = Article
\bibitem[{Gallino and Moreno(2014)}]{Gallino_Moreno2014}
\bibinfo{author}{Gallino, S.}, \bibinfo{author}{Moreno, A.},
  \bibinfo{year}{2014}.
\newblock \bibinfo{title}{Integration of online and offline channels in retail:
  The impact of sharing reliable inventory availability information}.
\newblock \bibinfo{journal}{Management Science} \bibinfo{volume}{60},
  \bibinfo{pages}{1434--1451}.
%Type = Article
\bibitem[{Gao and Su(2016)}]{Gao_Su2016}
\bibinfo{author}{Gao, F.}, \bibinfo{author}{Su, X.}, \bibinfo{year}{2016}.
\newblock \bibinfo{title}{Omnichannel retail operations with
  buy-online-and-pick-up-in-store}.
\newblock \bibinfo{journal}{Management Science} \bibinfo{volume}{63},
  \bibinfo{pages}{2478--2492}.
%Type = Article
\bibitem[{Harsha et~al.(2019)Harsha, Subramanian and
  Uichanco}]{Harsha_Subramanian_Uichanco2019}
\bibinfo{author}{Harsha, P.}, \bibinfo{author}{Subramanian, S.},
  \bibinfo{author}{Uichanco, J.}, \bibinfo{year}{2019}.
\newblock \bibinfo{title}{Dynamic pricing of omnichannel inventories: Honorable
  mention—2017 {M}\&{SOM} practice-based research competition}.
\newblock \bibinfo{journal}{Manufacturing \& Service Operations Management}
  \bibinfo{volume}{21}, \bibinfo{pages}{47--65}.
%Type = Article
\bibitem[{H{\"u}bner et~al.(2016)H{\"u}bner, Holzapfel and
  Kuhn}]{Hubner_Holzapfel_Kuhn2016}
\bibinfo{author}{H{\"u}bner, A.}, \bibinfo{author}{Holzapfel, A.},
  \bibinfo{author}{Kuhn, H.}, \bibinfo{year}{2016}.
\newblock \bibinfo{title}{Distribution systems in omni-channel retailing}.
\newblock \bibinfo{journal}{Business Research} \bibinfo{volume}{9},
  \bibinfo{pages}{255--296}.
%Type = Book
\bibitem[{Ishikawa and Nejo(2002)}]{Ishikawa_Nejo2002}
\bibinfo{author}{Ishikawa, A.}, \bibinfo{author}{Nejo, T.},
  \bibinfo{year}{2002}.
\newblock \bibinfo{title}{Success of 7-eleven Japan: Discovering the Secrets of
  the World's Best-Run Convenience Chain Stores}.
\newblock \bibinfo{publisher}{Singapore: World Scientific}.
%Type = Article
\bibitem[{Li et~al.(2005)Li, Sun and Wilcox}]{Li_Sun_Wilcox2005}
\bibinfo{author}{Li, S.}, \bibinfo{author}{Sun, B.}, \bibinfo{author}{Wilcox,
  R.T.}, \bibinfo{year}{2005}.
\newblock \bibinfo{title}{Cross-selling sequentially ordered products: An
  application to consumer banking services}.
\newblock \bibinfo{journal}{Journal of Marketing Research}
  \bibinfo{volume}{42}, \bibinfo{pages}{233--239}.
%Type = Article
\bibitem[{Netessine and Rudi(2006)}]{Netessine_Rudi2006}
\bibinfo{author}{Netessine, S.}, \bibinfo{author}{Rudi, N.},
  \bibinfo{year}{2006}.
\newblock \bibinfo{title}{Supply chain choice on the internet}.
\newblock \bibinfo{journal}{Management Science} \bibinfo{volume}{52},
  \bibinfo{pages}{844--864}.
%Type = Article
\bibitem[{Netessine et~al.(2006)Netessine, Savin and
  Xiao}]{Netessine_Savin_Xiao2006}
\bibinfo{author}{Netessine, S.}, \bibinfo{author}{Savin, S.},
  \bibinfo{author}{Xiao, W.}, \bibinfo{year}{2006}.
\newblock \bibinfo{title}{Revenue management through dynamic cross selling in
  e-commerce retailing}.
\newblock \bibinfo{journal}{Operations Research} \bibinfo{volume}{54},
  \bibinfo{pages}{893--913}.
%Type = Article
\bibitem[{Ofek et~al.(2011)Ofek, Katona and Sarvary}]{Ofek_Katona_Sarvary2011}
\bibinfo{author}{Ofek, E.}, \bibinfo{author}{Katona, Z.},
  \bibinfo{author}{Sarvary, M.}, \bibinfo{year}{2011}.
\newblock \bibinfo{title}{“{B}ricks and clicks”: The impact of product
  returns on the strategies of multichannel retailers}.
\newblock \bibinfo{journal}{Marketing Science} \bibinfo{volume}{30},
  \bibinfo{pages}{42--60}.
%Type = Article
\bibitem[{Rigby(2011)}]{Rigby2011}
\bibinfo{author}{Rigby, D.}, \bibinfo{year}{2011}.
\newblock \bibinfo{title}{The future of shopping}.
\newblock \bibinfo{journal}{Harvard Business Review} \bibinfo{volume}{89},
  \bibinfo{pages}{65--76}.
%Type = Article
\bibitem[{Rosenblum and Kilcourse(2013)}]{Rosenblum_Kilcourse2013}
\bibinfo{author}{Rosenblum, P.}, \bibinfo{author}{Kilcourse, B.},
  \bibinfo{year}{2013}.
\newblock \bibinfo{title}{Omni-channel 2013: The long road to adoption}.
\newblock \bibinfo{journal}{Benchmark Report, RSR Research, Miami} .
%Type = Article
\bibitem[{Stokey(1981)}]{Stokey1981}
\bibinfo{author}{Stokey, N.L.}, \bibinfo{year}{1981}.
\newblock \bibinfo{title}{Rational expectations and durable goods pricing}.
\newblock \bibinfo{journal}{The Bell Journal of Economics}
  \bibinfo{volume}{12}, \bibinfo{pages}{112--128}.
%Type = Article
\bibitem[{Su and Zhang(2009)}]{Su_Zhang2009}
\bibinfo{author}{Su, X.}, \bibinfo{author}{Zhang, F.}, \bibinfo{year}{2009}.
\newblock \bibinfo{title}{On the value of commitment and availability
  guarantees when selling to strategic consumers}.
\newblock \bibinfo{journal}{Management Science} \bibinfo{volume}{55},
  \bibinfo{pages}{713--726}.
%Type = Article
\bibitem[{Terasaka(1998)}]{Terasawa1998}
\bibinfo{author}{Terasaka, A.}, \bibinfo{year}{1998}.
\newblock \bibinfo{title}{Development of new store types: The role of
  convenience stores in {J}apan}.
\newblock \bibinfo{journal}{GeoJournal} \bibinfo{volume}{45},
  \bibinfo{pages}{317--325}.

\end{thebibliography}
\clearpage
\par
\noindent
Table 1: \emph{Results of Equilibrium with respect to $q$}
\par
\vspace{\baselineskip}
\begin{threeparttable}
\scalebox{0.9}{
%\centering
\begin{tabular}{p{330pt}p{230pt}}
  \hspace{120pt} $0 \leq \bar{\mu} \leq c_{o}$
 &
  \hspace{40pt} $c_{o} < \bar{\mu} \leq M$
\\
\\ \hline
\\
  \textbf{BOPS I ($\mathbf{\xi = 1}$):}\ \
  $\bigl( \frac{c}{p - c} \bigr)c_{o} \leq \bar{\mu} \leq c_{o}$,
 &
  \ \
\\
\\
  \hspace{50pt} $q = c_{o}$,\ \ $\pi = (p - c)c_{o} - C(\bar{\mu})$
 &
  \textbf{Store I ($\mathbf{\xi = 1}$):}\ \
  $p \geq 2c$,
  \ \
\\
\\
  \textbf{BOPS II ($\mathbf{0 < \xi < 1}$):}\ \
  $\bigl( \frac{c}{p + c} \bigr)c_{o} \leq \bar{\mu} \leq \bigl( \frac{c}{p - c} \bigr)c_{o}$,
 &
  \hspace{50pt} $q = c_{o}$,
  \ \
\\
\\
  \hspace{20pt} $q = \frac{1}{4\bar{\mu}}\Bigl[ \left( \frac{p\bar{\mu}}{c} \right)^{2} - (c_{o} - \bar{\mu})^{2} \Bigr]$,\ \ $\pi = \frac{c}{4\bar{\mu}}\Bigl[ \left( \frac{p - c}{c} \right)\bar{\mu} + c_{o} \Bigr]^{2} - C(\bar{\mu})$
 &
  \hspace{30pt} $\pi = (p - c)c_{o} - C(\bar{\mu})$
  \ \
\\
\\
  \textbf{BOPS III ($\mathbf{\xi = 0}$):}\ \
  $0 \leq \bar{\mu} < \bigl( \frac{c}{p + c} \bigr)c_{o}$,
 &
  \ \
\\
\\
  \hspace{50pt} $q = 0$,\ \ $\pi = p(c_{o} - \bar{\mu}) - C(\bar{\mu})$
 &
  \ \
\\
\\ \hline
\\
  \textbf{BOPS II ($\mathbf{0 < \xi < 1}$):}\ \
  $\bigl( \frac{c}{p + c} \bigr)c_{o} \leq \bar{\mu} \leq \bigl( \frac{c}{p - c} \bigr)c_{o}$,
 &
  \textbf{Store II ($\mathbf{0 < \xi < 1}$):}\ \ $c < p < 2c$,
  \ \
\\
\\
  \hspace{20pt} $q = \frac{1}{4\bar{\mu}}\Bigl[ \left( \frac{p\bar{\mu}}{c} \right)^{2} - (c_{o} - \bar{\mu})^{2} \Bigr]$,\ \ $\pi = \frac{c}{4\bar{\mu}}\Bigl[ \left( \frac{p - c}{c} \right)\bar{\mu} + c_{o} \Bigr]^{2} - C(\bar{\mu})$
 &
  \hspace{50pt} $q = \left( \frac{p}{2c} \right)^{2}c_{o}$,
  \ \
\\
\\
  \textbf{BOPS III ($\mathbf{\xi = 0}$):}\ \
  $0 \leq \bar{\mu} < \bigl( \frac{c}{p + c} \bigr)c_{o}$,
 &
  \hspace{30pt} $\pi = \left( \frac{p^{2}}{4c} \right)c_{o} - C(\bar{\mu})$
  \ \
\\
\\
  \hspace{50pt} $q = 0$,\ \ $\pi = p(c_{o} - \bar{\mu}) - C(\bar{\mu})$
 &
  \ \
\\
\\ \hline
\end{tabular}
}% end of scalebox
\vspace{\baselineskip}
\begin{tablenotes}[para,flushleft,online,normal]
\emph{Note.}\ Upper side: $p\geq 2c$; lower side: $c < p < 2c$.
\end{tablenotes}
\end{threeparttable}
\par
\vspace{5\baselineskip}
\noindent
Table 2: \emph{Results of Simulation}
\par
\vspace{\baselineskip}
\begin{threeparttable}
\scalebox{0.9}{
\begin{tabular}{p{2.0cm}cccccccccccccccccccc}
($r = 0.1$) \\
  \hline
  Week & 0 & 1 & 2 & 3 & 4 & 5 & 6 & 7 & 8 & 9 & 10 & 11 & 12 & 13 & 14 & 15 & 16 & 17 & 18 & 19 \\
  \hline
  Store 0 & S & B & B & S & S & S & S & S & S & B & S & S & S & S & S & S & S & S & S & S \\
  \hline
  Store 1 & S & S & S & S & S & S & S & B & S & S & S & B & S & S & S & S & S & S & S & S \\
  \hline
\\ \\
($r = 0.3$) \\
  \hline
  Week & 0 & 1 & 2 & 3 & 4 & 5 & 6 & 7 & 8 & 9 & 10 & 11 & 12 & 13 & 14 & 15 & 16 & 17 & 18 & 19 \\
  \hline
  Store 0 & S & B & B & S & S & S & S & S & S & B & B & S & S & S & B & S & B & S & B & S \\
  \hline
  Store 1 & S & S & S & S & S & B & S & B & B & S & S & B & B & S & S & S & B & S & S & S \\
  \hline
\\ \\
($r = 0.5$) \\
  \hline
  Week & 0 & 1 & 2 & 3 & 4 & 5 & 6 & 7 & 8 & 9 & 10 & 11 & 12 & 13 & 14 & 15 & 16 & 17 & 18 & 19 \\
  \hline
  Store 0 & S & B & B & B & B & B & S & S & S & B & B & B & S & S & B & S & B & S & B & S \\
  \hline
  Store 1 & S & B & S & S & S & B & S & B & B & S & S & B & B & B & B & S & B & S & S & S \\
  \hline
\end{tabular}
}% end of scalebox
\vspace{\baselineskip}
\begin{tablenotes}[para,flushleft,online,normal]
\emph{Note.}\ $S$ and $B$ denote Store I and BOPS III, respectively.
Both stores adopt Store I in the first week.
\end{tablenotes}
\end{threeparttable}
\end{document}